\documentclass{article}[fontsize=12pt]
\usepackage[margin=1in]{geometry}
\usepackage{natbib}
\usepackage{enumerate}
\usepackage{amsmath}               
\usepackage{amsfonts}
\usepackage{amssymb}
\usepackage{amsthm}
\usepackage{graphicx}
\usepackage{color,soul}
\usepackage{tikz}
\usepackage{subcaption}
\usepackage[T1]{fontenc}
\usepackage{lmodern}
\usepackage{tikz}
\usepackage{bm}
\usepackage{bbm}
\usepackage{multirow}
\usepackage{hyperref}

\usetikzlibrary{shapes,backgrounds,fit}

\setcitestyle{square,numbers}

\usepackage{array,ragged2e}
\newcolumntype{P}[1]{>{\RaggedRight\arraybackslash}p{#1}}

\usetikzlibrary{positioning}

\theoremstyle{plain}
\newtheorem{assumption}{Assumption}

\newtheorem{lemma}{Lemma}

\providecommand{\keywords}[1]{\textit{Keywords}: #1}
\begin{document}
\title{Sequential linear regression for conditional mean imputation of longitudinal continuous outcomes under reference-based assumptions}
\date{}
      
\maketitle
\vspace{-2cm}
{\center{Sean Yiu\\
       Roche Products Limited, Welwyn Garden City, UK.\\
      Email: sean.yiu@roche.com\\}}
      
\begin{abstract}
In clinical trials of longitudinal continuous outcomes, reference based imputation (RBI) has commonly been applied to handle missing outcome data in settings where the estimand incorporates the effects of intercurrent events, e.g. treatment discontinuation. RBI was originally developed in the multiple imputation framework, however recently conditional mean imputation (CMI) combined with the jackknife estimator of the standard error was proposed as a way to obtain deterministic treatment effect estimates and correct frequentist inference. For both multiple and CMI, a mixed model for repeated measures (MMRM) is often used for the imputation model, but this can be computationally intensive to fit to multiple data sets (e.g. the jackknife samples) and lead to convergence issues with complex MMRM models with many parameters. Therefore, a step-wise approach based on sequential linear regression (SLR) of the outcomes at each visit was developed for the imputation model in the multiple imputation framework, but similar developments in the CMI framework are lacking. In this article, we fill this gap in the literature by proposing a SLR approach to implement RBI in the CMI framework, and justify its validity using theoretical results and simulations. We also illustrate our proposal on a real data application.              
\end{abstract}

\section{Introduction}
The introduction of the ICH E9 addendum on estimands and sensitivity analysis in clinical trials \cite{ICH} has bought much needed clarity to the description of estimands for treatment effects in clinical trials. A main component of the addendum is the need to precisely specify how intercurrent events (IE) are handled, i.e. events occurring after treatment initiation that affect either the interpretation or the existence of the measurements associated with the clinical question of interest. In general, the IE handling strategy should primarily be chosen to align with the study objectives, however it has been noted that some strategies can be more robust with respect to missing data than others \cite{Mallinckrodt2019}. Discontinuation of the randomized treatment is an IE that commonly occurs in clinical trials, and is often handled using either the hypothetical or treatment policy strategy. The hypothetical strategy excludes the effect of treatment discontinuation in the estimand as it requires contrasting counterfactual outcomes from the setting where treatment discontinuation has been prevented. In contrast, the treatment policy strategy includes the effect of treatment discontinuation as it instead requires contrasting observed outcomes irrespective of the patient's treatment discontinuation status. See \cite{Nocci2022} for an illustration on the use and interpretation of the hypothetical and treatment policy strategies for handling treatment discontinuation in early Parkinson's disease. The focus of this article is on the setting where treatment discontinuation is the only noteworthy IE, and it is of interest to use the treatment policy strategy to handle treatment discontinuation when contrasting longitudinal continuous outcomes in randomized clinical trials.

Unfortunately, missing data in clinical trials are ubiquitous. Missing data may arise from IE, e.g. patients that discontinue treatment may be at higher risk of withdrawing from the study, or due to other reasons, e.g. administrative reasons. A popular approach to handle missing data is to apply missing data imputation, which involves replacing missing values with known values based on other available information. See Chapter 25 in \cite{Gelman2006} for a very accessible introduction into a variety of missing data imputation techniques and how they can be implemented in practice, and \cite{Lin2020} for a comprehensive review of missing data imputation techniques focused on the experimental design setting. For treatment policy estimation, it is imperative that the chosen approach to handle missing data can capture the effect of IE on subsequent outcomes \cite{Wang2023}. For instance, if there are sufficient observed data post treatment discontinuation, then such data can be used to impute missing data post treatment discontinuation, e.g. see \cite{Drury2023} for a proposal. Reference based imputation (RBI) methods for treatment policy estimation were proposed for the situation where observed data post treatment discontinuation are sparse but a clinically plausible assumption on the impact of treatment discontinuation on subsequent outcomes can be made \cite{Carpenter2013}. Specifically, RBI methods replace missing data post treatment discontinuation in the active arm using the postulated assumption on the impact of treatment discontinuation and the observed trend in the outcomes in a reference population, e.g. patients in the control arm. See \cite{Cro2020} for a tutorial on RBI methods and \cite{Wang2023} for recent example case studies on the use of RBI methods. Originally, RBI was developed in the multiple imputation framework where Rubin's combination rules are used for inference. Under this framework, it has be shown that inferences are approximately information anchored \cite{Cro2019}, i.e. the use of a reference population to impute missing data does not decrease the proportion of information lost due to missing data. Although the notion of information anchored inference is logical, many researchers (e.g. see \cite{Seaman2014} and \cite{Bartlett2023}) have observed that such inferences are often conservative, e.g. confidence intervals have higher coverage probabilities than the nominal value, which could be undesirable for the primary estimator that determines the overall conclusions of the trial. Subsequently, researchers have considered alternative approaches by either deriving analytical expressions for the correct frequentist asymptotic variance (e.g. see \cite{Lu2014} and \cite{Tang2017}), or proposed an approximation by applying multiple imputation in bootstrap samples of the observed data \cite{Vonhippel2021}.  

Recently, \cite{Wolbers2022} developed RBI in the conditional mean imputation (CMI) framework and proposed to use the jackknife for inference. This approach leads to deterministic treatment effect estimates (i.e. estimates that are unaffected by Monte Carlo error) and correct frequentist inference. A limitation of this approach is that the jackknife requires repeated fitting of mixed models for repeated measures models (MMRM) which is computationally intensive and may lead to convergence problems for complex MMRM models with many parameters. In order to decrease computational cost in the multiple imputation framework, several researchers have proposed a step-wise approach consisting of only applying a sequence of linear regressions (SLR) to estimate the parameters of the imputation model (see the presentation by Michael O’Kelly and James Roger in March 2013 on the \href{https://www.lshtm.ac.uk/research/centres-projects-groups/missing-data#dia-missing-data}{Drug Information Association Scientific Working Group on Estimands and Missing Data website} and e.g. \cite{Fang2022} for more details). In this article, we adapt this idea to develop a new step-wise approach based on SLR to implement RBI in the CMI framework, justify its validity using theoretical results and simulations, and provide an illustration to a real data application. Furthermore, a short tutorial on how to implement the proposal is included in the Supplementary Materials. 

The rest of this article is organized as follows. In Sections 2 and 3, we introduce the notation, setting of interest and assumptions for making inference. In Section 4, we provide high-level details of the recently proposed CMI with MMRM approach. In Sections 5 and 6, we describe and justify the proposed estimator. In Section 7, we propose methods for inference. In Section 8, we highlight critical differences between the implementation of MMRM and SLR. In Sections 9 and 10, we compare the proposed method to its MMRM counterpart in simulations and a real data application. Finally, we conclude with a discussion in Section 11.

\section{Setting}
We consider the setting where patients are randomized to either an active treatment or control arm at baseline (denoted by visit $j=0$), where a set of baseline covariates $\bm{X}$ is always completely recorded, but the outcome at baseline $Y_{0}$ may be incomplete. The patients are then followed-up over time at a common set of visit times $j=1,\ldots,M$. Let the random variable $Z$ represent the arm to which the patient was randomized, with $Z=1$ representing assignment to the active arm and $Z=0$ to the control arm. At each visit $j$ ($j=0,\ldots,M$), including at baseline, let $Y_j$ be the continuous outcome, $R_{j}$ be the missing data indicator specifying whether $Y_j$ was observed ($R_j=1$) or unobserved ($R_j=0$) and $Y^\text{obs}_j=Y_j$ if $R_j=1$, otherwise $Y^\text{obs}_j$ is unavailable. For convenience, let the history of a variable up to and including visit $j$ be denoted by an overline, e.g. $\overline{\bm{Y}}_j=\{Y_0,\ldots,Y_j\}$. In our setting, patient outcomes can be unobserved at any visit and non-monotone missing patterns may arise, i.e $R_j=0$ does not imply $R_{j+1}=0$. Next, let the random variable $D$ be the first visit where the patient did not receive their assigned treatment. In our setting, the possible values of $D$ are $0,\ldots,M+1$, where $D=0$ means that the patient never received their assigned treatment at any time during the study and $D=M+1$ means that the patient received treatment at baseline and all $M$ post-baseline visits, i.e. no treatment discontinuation occurred. We assume that a reference-based assumption for post treatment discontinuation outcomes is clinically plausible and that only discontinuing from the active treatment is expected to have a profound impact on the outcomes; see Section~\ref{Assumptions} for more details. For instance, this setting may arise if patients who discontinue from active or control treatment cannot subsequently initiate more efficacious treatments than the control treatment, e.g. due to a lack of availability. Furthermore, although we allow for the possibility that patients can still contribute outcome data after treatment discontinuation, we assume that such data are likely to be sparse. 
 
Motivated by the causal framework of \cite{White2019}, we define the potential outcome $Y_j(d,z)$ to be the outcome that would have been observed at visit $j$ had the patient received treatment $z$ from baseline until visit $d-1$ but not thereafter. For instance, $Y_j(M+1,1)$ and $Y_j(M+1,0)$ are potential outcomes at visit $j$ if the patients never discontinued the active and control treatment up to visit $j$, respectively. The potential outcomes are linked to the observed outcomes via $Y_j=Y_j(D,Z)$ (the consistency assumption), but are otherwise unobserved. Two estimands that are often of interest at a particular visit $j$ are $E[Y_j(M+1,1)-Y_j(M+1,0)]$ (hypothetical estimand) and $E[Y_j(D,1)-Y_j(D,0)]$ (treatment policy estimand), which are constructed by applying the hypothetical and treatment policy strategies to handle treatment discontinuation. The hypothetical estimand at visit $j$ is the mean difference in outcomes at visit $j$ had all patients always received the active versus control treatment up to visit $j$. Hence, the hypothetical estimand excludes the effect of treatment discontinuation on the outcomes. In contrast, the treatment policy estimand at visit $j$ is the mean difference in outcomes at visit $j$ under the observed treatment trajectories had all patients been assigned the active versus control treatment. Thus, the treatment policy estimand includes the effect of treatment discontinuation on the outcomes, unlike the hypothetical estimand. In this article, our primary focus will be on estimating the treatment policy estimand.

\section{Assumptions}\label{Assumptions}
We make the following assumptions to identify the treatment policy estimand of interest:
\begin{assumption}[Ignorability of treatment assignment]
$E[Y_j(d,z) \mid Z]=E[Y_j(d,z)]$ \text{for all} $d$ and $z$.
\end{assumption}
This assumption is plausible in our setting because treatment has been randomly assigned. Under this assumption, $E[Y_j(D,z)]=E[Y_j(D,z) \mid Z=z]=E[Y_j \mid Z=z]$, where the last equality follows from the consistency assumption. Thus, the treatment policy estimand is equal to $E[Y_j\mid Z=1]-E[Y_j\mid Z=0]$. In observational settings, this assumption can be further relaxed by additionally conditioning on $\bm{X}$, i.e. $E[Y_j(d,z) \mid Z, \bm{X}]=E[Y_j(d,z)| \bm{X}]$.

\begin{assumption}[Missing at random (MAR) conditional on assignment to control]
$E[Y_j \mid \overline{\bm{Y}}^\text{obs}_{j-1}, \bm{X}, Z=0,D,R_j=1]=E[Y_j \mid \overline{\bm{Y}}^\text{obs}_{j-1}, \bm{X}, Z=0,D]=E[Y_j \mid \overline{\bm{Y}}^\text{obs}_{j-1}, \bm{X}, Z=0]$~\text{for}~$j=0,\ldots,M$,~\text{where} $Y_{-1}=\emptyset$.
\end{assumption}

The first equality of this assumption states that mean trajectories of the outcome at each visit in the control arm can be estimated from the complete data within levels of the previously observed outcomes, treatment discontinuation status and baseline covariates. The second equality states that discontinuation of the control treatment has no impact on the outcomes. This assumption is likely to be most plausible when the control treatment is placebo, but could also be plausible if patients can initiate other standard of care treatments with similar efficacy to the control treatment after discontinuation from the control treatment. In theory, this assumption could be relaxed by also accounting for the impact of treatment discontinuation on the outcomes (e.g. see \cite{Drury2023}, \cite{Guizzaro2021}). However, in practice it is difficult to characterize the effect of treatment discontinuation on the outcomes when data after treatment discontinuation are sparse.

In regards to the assumption on the ignorability of the missing data indicator, we follow \cite{Robins1997} and acknowledge that it is difficult to conceive of non-monotone MAR mechanisms that are not also missing completely at random (MCAR), i.e. where ignorability of the missing data indicator is achieved conditional on at most only the completely observed baseline covariates $\bm{X}$. If however it is possible to restrict the missing data patterns to be monotone, then $\overline{\bm{Y}}^\text{obs}_{j-1}=\overline{\bm{Y}}_{j-1}$ when $R_j=1$ in Assumption 2, which leads to more conceivable missing data generating mechanisms under Assumption 2 that are not MCAR.

\begin{assumption}[MAR prior to treatment discontinuation and conditional on assignment to active treatment]
 $E[Y_j \mid \overline{\bm{Y}}^\text{obs}_{j-1}, \bm{X}, Z=1,D > j,R_j=1]=E[Y_j \mid \overline{\bm{Y}}^{\text{obs}}_{j-1}, \bm{X}, Z=1, D > j]$~\text{for}~$j=0,\ldots,M$,~\text{where} $Y_{-1}=\emptyset$.   
\end{assumption}
   Similarly to Assumption 2, this assumption allows us to estimate mean trajectories of outcomes prior to treatment discontinuation in the active arm using only the previously observed outcomes and baseline covariates. For outcomes post treatment discontinuation, a separate reference based assumption will be made to describe the behavior of their mean trajectories; see Assumption 5 for more details.
   
   Next, we will describe the exact form of the outcomes conditional on previous outcomes and baseline covariates under the control treatment and up to the time of discontinuation from the active treatment. 
\begin{assumption}[The conditional expectation of always treated potential outcomes follow a linear model]
 $E[Y_j(M+1,z) \mid \overline{\bm{Y}}_{j-1}(M+1,z), \bm{X},D]=\bm{\beta}^t_j(z)\cdot\overline{\bm{Y}}_{j-1}(M+1,z)+\bm{\beta}^{'t}_j(z)\cdot(1,\bm{X})$ for $j=0,\ldots,M$, where $Y_{-1}(M+1,z)=\emptyset$ for all $k$. 
\end{assumption}
This assumption would be satisfied if $\overline{\bm{Y}}_{j}(M+1,z) \mid \bm{X}$ follows a homoscedastic multivariate normal distribution, where each component of the mean vector is a linear combination of $\bm{X}$. For instance, if $\overline{\bm{Y}}_j(M+1,z)$ has a multivariate normal distribution with mean vector $\overline{\bm{\mu}}_j(z)$ and covariance matrix $\Sigma_j(z)$, then $E[Y_j(M+1,z) \mid \overline{\bm{Y}}_{j-1}(M+1,z)]=\mu_j(z)+\Sigma_{j,1:(j-1)}(z)\Sigma^{-1}_{1:(j-1),1:(j-1)}(z)\cdot\{\overline{\bm{Y}}_{j-1}(z)-\overline{\bm{\mu}}_{j-1}(z)\}$, where $\Sigma_{a_1:a_2,b_1:b_2}(z)$ is the sub matrix of $\Sigma(z)$ formed by selecting the $a_1,\ldots,a_2$th rows and $b_1,\ldots,b_2$th columns of $\Sigma(z)$, $a_1:a_1=a_1$, and $\bm{a}\cdot \bm{b}$ denotes the dot product between the vectors $\bm{a}$ and $\bm{b}$.

In Section~\ref{Jus_sec}, it will be useful to realize that Assumption 4 implies $E[Y_j \mid \overline{\bm{Y}}_{j-1}, \bm{X},Z=z,D > j]=E[Y_j(M+1,z) \mid \overline{\bm{Y}}_{j-1}(M+1,z),\bm{X},D > j]=\bm{\beta}^t_j(z)\cdot\overline{\bm{Y}}_{j-1}+\bm{\beta}^{'t}_j(z)\cdot(1,\bm{X})$ when $D >j$.
   
\begin{assumption}[Reference-based imputation] For visits $j\geq k$ ($j=0,\ldots,M$), the marginal mean of the outcomes in the active arm after treatment discontinuation is 
\[ 
E[Y_j \mid \bm{X}, Z=1, D=k]=
\begin{cases}
    \mu_j(0),& \text{Jump to Reference (J2R)}\\
    \{\mu_{k-1}(1)-\mu_{k-1}(0)\}+\mu_j(0),              & \text{Copy Increments in Reference (CIR)}
\end{cases}
\]
where $\mu_{j}(1)=E[Y_j(M+1,1)\mid \bm{X}]$ and $\mu_{j}(0)=E[Y_j(M+1,0)\mid \bm{X}]$, i.e. the mean outcomes had the patients always received the active and control treatment. Whereas the conditional (on the outcome history) mean of the outcomes in the active arm after treatment discontinuation is
\[ 
E[Y_j \mid \overline{\bm{Y}}_{j-1}, \bm{X}, Z=1, D=k]=
\begin{cases}
    \bm{\beta}^t_{j}(0)\cdot \{\overline{\bm{Y}}_{j-1}-\bm{\mu}^\text{J2F}_{j-1}\}+\mu_j(0),& J2R\\
    \bm{\beta}^t_{j}(0)\cdot\{\overline{\bm{Y}}_{j-1}-\bm{\mu}^\text{CIR}_{j-1}\}+\{\mu_{k-1}(1)-\mu_{k-1}(0)\}+\mu_j(0),              & CIR
\end{cases}
\]
where $\mu^\text{J2F}_{j-1}=[\mu_0(1),\ldots,\mu_{k-1}(1),\mu_{k}(0),\ldots,\mu_{j-1}(0)]$ and $\mu^\text{CIR}_{j-1}=[\mu_0(1),\ldots,\mu_{k-1}(1),\{\mu_{k-1}(1)-\mu_{k-1}(0)\}+\mu_{k}(0),\ldots,\{\mu_{k-1}(1)-\mu_{k-1}(0)\}+\mu_{j-1}(0)]$.  
\end{assumption}
In words, under J2F the marginal mean of the outcomes (unconditional on the outcome history) at visit $j$ in the subgroup who discontinued from active treatment at visit $k \leq j$ is equal to the marginal mean of the outcomes in the control arm at visit $j$, within levels of the baseline covariates. In contrast, CIR states that the same subgroup marginal mean is the treatment effect based on the hypothetical estimand at visit $k-1$ plus the marginal mean of the outcomes in the control arm at visit $j$, within levels of the baseline covariates. Therefore, unlike J2R, CIR attempts to encode the benefit of having previously been exposed to the active treatment.

It is worth noting that the J2F and CIR assumptions stated in Assumption 5 are based on those in \cite{Wolbers2022}, which extend the J2F and CIR assumptions in \cite{Carpenter2013} to the scenario where off-treatment data are observed. Recently, \cite{Liu2022} described a different J2R assumption that is also applicable when off-treatment data are observed. Their assumption states that $E[Y_j \mid \overline{\bm{Y}}_{j-1}, \bm{X}, Z=1, D \leq j]=E[Y_j \mid \overline{\bm{Y}}_{j-1}, \bm{X}, Z=0]$. After taking expectation over the outcome history, the marginal mean as expressed in our notation is $E[Y_j \mid \bm{X}, Z=1, D=k]=\bm{\beta}_{k,j}(0) \cdot \{\mu_{k-1}(1)-\mu_{k-1}(0)\}+\mu_j(0)$, where $\bm{\beta}_{k,j}(0)$ is the vector of regression coefficients for $\overline{\bm{Y}}_{k-1}$ in the model $E[Y_j\mid \overline{\bm{Y}}_{k-1},\bm{X},Z=0,D>j]$. Hence, the J2R assumption in \cite{Liu2022} is different to the J2R and CIR assumptions stated in Assumption 5. Moreover, depending on the correlation between $Y_j$ and $\overline{\bm{Y}}_{k-1}$, it can result in less conservative estimates of the treatment policy estimand than assuming CIR or more conservatives estimates than assuming J2R in Assumption 5.

\section{CMI based on MMRM}\label{MMRM_sec}
Recently, \cite{Wolbers2022} developed a consistent estimator of the treatment policy estimand under the assumptions stated in Section~\ref{Assumptions}. Their estimator, which we will refer to as CMI with MMRM, first requires estimating $\overline{\bm{\mu}}_M(0)$, $\overline{\bm{\mu}}_M(1)$ and $\bm{\beta}^t_{j}(0)$ ($j=1,\ldots,M$) as defined in Assumption 5, by fitting a MMRM with unstructured covariance matrix to the outcomes in the control and active treatment arm up to the last visit prior to treatment discontinuation. CMI from the fitted MMRM are then used to impute missing data under Assumptions 2, 3 and 5 in Section~\ref{Assumptions}. Finally, the treatment policy estimand at visit $j$ is estimated by fitting a simple linear model to $Y_j$ with $Z$ and possibly a subset of $X$ as covariates to the outcome data after imputation. Regarding inference, \cite{Wolbers2022} proposed to use resampling based methods, e.g. the jackknife or bootstrap. The entire process from estimating the MMRM for the imputation models to making inference on the estimands can be implemented in the rbmi package \cite{Gower-Page2022} in R \cite{R2022}. The main advantages of this estimator are that it delivers deterministic treatment effect estimates and correct frequentist inference. However, the potential disadvantages are that it can be computationally intensive and may result in convergence issues when the imputation model contain many parameters because MMRM need to be fitted to many permutations of the data \cite{Mallinckrodt2008}.         

In the next section, we describe a new estimator of the treatment policy estimand that does not involve fitting MMRM to estimate $\overline{\bm{\mu}}_M(1)$, $\overline{\bm{\mu}}_M(0)$ and $\bm{\beta}^t_{j}(0)$ ($j=0,\ldots,M$), but instead a sequence of linear models for outcomes at each visit. We then explain in Section~\ref{MMRMvsSLR} why our proposed approach can be much more computationally efficient and at less risk of convergence issues compared to the CMI with MMRM approach.

\section{Proposed estimator}\label{Prop_est}
Let $Y^H_{i,0}=R_{i,0}Y_{i,0}+(1-R_{i,0})\hat{\eta}_{i,0}(Z_i)$, where the subscript $i$ ($i=1,\ldots,n$) represents the $i$th patient, e.g. $Y_{i,j}$ represents the observed outcome from the $i$th patient at visit $j$, $\hat{\eta}_{i,0}(Z_i)=\hat{\bm{\beta}}^{'}_0(Z_i)\cdot (1,\bm{X}_i)$ and $\hat{\bm{\beta}}^{'}_0(z)$ ($z=0,1$) are the estimated regression coefficients of $(1,\bm{X}_i)$ in the linear model for $Y_{i,0}$ conditional on $\bm{X}_{i}$ in patients with $R_{i,0}=1$ and $Z_i=z$. Next, define sequentially for $j=1,\ldots,M$, 
\[
Y^H_{i,j} = 
  R_{i,j}\mathbbm{1}(D_i> j)Y_{i,j}+\{1-R_{i,j}\mathbbm{1}(D_i> j)\}\hat{\eta}_{i,j}(z)
\]
where $\hat{\eta}_{i,j}(Z_i)=\{\hat{\bm{\beta}}^{'}_j(Z_i),\hat{\bm{\beta}}_j(Z_i)\}\cdot(1,\bm{X}_i,\overline{\bm{Y}}^H_{i,j-1})$, and $\hat{\bm{\beta}}^{'}_j(z)$ and $\hat{\bm{\beta}}_j(z)$ ($z=0,1$) are the estimated regression coefficients of $(1,\bm{X}_i)$ and $\overline{\bm{Y}}^H_{i,j-1}$ in the linear model for $Y_{i,j}$ conditional on $\bm{X}_{i}$ and $\overline{\bm{Y}}^H_{i,j-1}$ in patients with $Z_i=z$, $R_{i,j}=1$ and $D_i> j$. In other words, $Y^H_{i,j}$ is the observed outcome if the patient is still under their assigned treatment or a CMI based on a linear model fitted to patients who are still under their assigned treatment. Although not of primary interest, we replicate the results in \cite{Aalen2010} by showing in the next section that the following is a consistent estimator of the hypothetical estimand at visit $j$:
$$\frac{1}{|Z|}\sum_{i=1}^nZ_iY^H_{i,j}-\frac{1}{n-|Z|}\sum_{i=1}^n(1-Z_i)Y^H_{i,j},$$
where $|Z|=\sum_{i=1}^nZ_i$, if the assumptions in Section~\ref{Assumptions} are satisfied. In order to describe our proposed estimator of the treatment policy estimand, we will need to introduce further notation. Let $Y^{J2F}_{i,0}=Y^{CIR}_{i,0}=Y^H_{i,0}$. Then, define sequentially for $j=1,\ldots,M$, 
\[
Y^{J2R}_{i,j} = \begin{cases}
  R_{i,j}Y_{i,j}+(1-R_{i,j})\left[\mathbbm{1}(D_i> j)\hat{\eta}_{i,j}(1)+\mathbbm{1}(D_i\leq j)\left\{\hat{\bm{\beta}}_{j}(0)\{\overline{\bm{Y}}^{J2F}_{i,j-1}-\hat{\bm{\mu}}^{J2F}_{i,j-1}\}+\hat{\mu}_{i,j}(0)\right\}\right]           & Z_i=1\\
  R_{i,j}Y_{i,j}+(1-R_{i,j})\{\hat{\bm{\beta}}^{'}_j(0),\hat{\bm{\beta}}_j(0)\}\cdot(1,\bm{X}_i,\overline{\bm{Y}}^{J2R}_{i,j-1})     & Z_i=0
         \end{cases}
\]
where $\hat{\bm{\mu}}^{J2F}_{i,j-1}=\{\hat{\mu}_{i,0}(1),\ldots,\hat{\mu}_{i,D_i-1}(1),\hat{\mu}_{i,D_i}(0),\ldots,\hat{\mu}_{i,j-1}(0)\}$, $\hat{\mu}_{i,j}(z)=\hat{\bm{\beta}}^{*}_j(z)\cdot (1,\bm{X}_i)$, and $\hat{\bm{\beta}}^{*}_j(z)$ ($z=0,1$) are the estimated regression coefficients of $(1,\bm{X}_i)$ in the linear model for $Y^H_{i,j}$ conditional on $\bm{X}_{i}$ in patients with $Z_i=z$. Similarly, define sequentially for $j=1,\ldots,M$,

\[
Y^{CIR}_{i,j} = \begin{cases}
  R_{i,j}Y_{i,j}+(1-R_{i,j})\Bigl[\mathbbm{1}(D_i> j)\hat{\eta}_{i,j}(1)+ \\ 
  \mathbbm{1}(D_i\leq j)\left\{\hat{\bm{\beta}}_{j}(0)\{\overline{\bm{Y}}^{CIR}_{i,j-1}-\hat{\bm{\mu}}^{CIR}_{i,j-1}\}+\{\hat{\mu}_{i,D_i-1}(1)-\hat{\mu}_{i,D_i-1}(0)\}+\hat{\mu}_{i,j}(0)\right\}\Bigr]           & Z_i=1\\
  Y^{J2R}_{i,j}     & Z_i=0
         \end{cases}
\]
where $\hat{\bm{\mu}}^{CIR}_{i,j-1}=[\hat{\mu}_{i,0}(1),\ldots,\hat{\mu}_{i,D_i-1}(1),\{\hat{\mu}_{i,D_i-1}(1)-\hat{\mu}_{i,D_i-1}(0)\}+\hat{\mu}_{i,D_i}(0),\ldots,\{\hat{\mu}_{i,D_i-1}(1)-\hat{\mu}_{i,D_i-1}(0)\}+\hat{\mu}_{i,j-1}(0)]$. That is, $Y^{J2R}_{i,j}$ only differs from $Y^{CIR}_{i,j}$ for patients with $Z_i=1$ and when $D_i\leq j$, i.e. for patients in the active arm and after discontinuation from active treatment. Our proposed estimator of the treatment policy estimand is
\begin{equation}\label{prop_est1} 
\begin{cases}
  \frac{1}{|Z|}\sum_{i=1}^nZ_iY^{J2F}_{i,j}-\frac{1}{n-|Z|}\sum_{i=1}^n(1-Z_i)Y^{J2F}_{i,j}, & J2R\\
  \frac{1}{|Z|}\sum_{i=1}^nZ_iY^{CIR}_{i,j}-\frac{1}{n-|Z|}\sum_{i=1}^n(1-Z_i)Y^{CIR}_{i,j}, & CIR.
\end{cases}
\end{equation}

It is worth noting that our estimators reflect key features of the hypothetical and treatment policy estimands. Specifically, our estimators of the hypothetical and treatment policy estimands are equivalent prior to treatment discontinuation, i.e. $Y^{J2R}_{i,j}=Y^{CIR}_{i,j}=Y^H_{i,j}$, in accordance with the definition of the estimands. In addition, our estimators of the treatment policy estimand attempts to capture the effects of treatment discontinuation, whereas our estimator of the hypothetical estimand does not. This is because $Y^{J2R}_{i,j}$ and $Y^{CIR}_{i,j}$ post treatment discontinuation are either set to be the observed outcome or a CMI based on previous outcomes, including those that are post treatment discontinuation. In contrast, $Y^H_{i,j}$ post treatment discontinuation is constructed without the use of any post treatment discontinuation data. 

Our estimators can also be easily extended to estimate conditional treatment effects. In the absence of missing data (including treatment discontinuation if the hypothetical estimand is of interest), conditional treatment effects can be estimated by fitting a linear model for $Y_{i,j}$ conditional on $\bm{V}_i$, where $\bm{V}_i$ is a vector containing an intercept, $Z_i$ and a subset of baseline covariates that are of interest to condition on. When missing outcome data is present, we propose to apply the same procedure but with $Y^{H}_{i,j}$, $Y^{J2F}_{i,j}$ or $Y^{CIR}_{i,j}$ in place of $Y_{i,j}$. For instance, the proposed estimator based on CIR is $(V^TV)^{-1}V^T\bm{Y}^{CIR}_j$, where the design matrix $V=(\bm{V}^T_{1}\ldots,\bm{V}^T_{n})$ and $\bm{Y}^{CIR}_j=(Y^{CIR}_{1,j},\ldots,Y^{CIR}_{n,j})$. Note that setting $\bm{V}_i=(1,Z_i)$ results in the estimator described in \eqref{prop_est1}.

\section{Justification for the proposed estimator}\label{Jus_sec}
In this section, we justify the validity of the proposed estimators. For conciseness, we will focus on the setting where data are generated under CIR since essentially the same arguments are required for the other settings.

Recall that the proposed estimator of the conditional treatment effect based on CIR is $(V^TV)^{-1}V^T\bm{Y}^{CIR}_j$. This estimator reduces to $(V^TV)^{-1}V^T\bm{Y}_j$ in the absence of missing outcome data, which we have assumed is unbiased. It then follows that if $E[Y^{CIRp}_{i,j}\mid \bm{V}_i]=E[Y_{ij}\mid \bm{V}_i]$, where $Y^{CIRp}_{i,j}$ is the probability limit of $Y^{CIR}_{i,j}$, the proposed estimator of the treatment effect will be asymptotically unbiased. 

For convenience, we will now show that $E[Y^{CIRp}_{i,j}\mid \bm{V}_i, Z_i=1]=E[Y_{i,j}\mid \bm{V}_i, Z_i=1]$, since the approach for the $Z_i=0$ case is analogous. This will require the following lemmas:

\begin{lemma}\label{Lem1}
$E[Y_{i,j}\mid \overline{\bm{Y}}^\text{obs}_{i,j-1},\bm{X}_i,Z_i=z,D_i>j]=\bm{\beta}^t_j(z)\cdot \overline{\bm{Y}}^{Ht}_{i,j-1}+\bm{\beta}^{'t}_j(z)\cdot (1,\bm{X}_i):=E[Y_{i,j}\mid \overline{\bm{Y}}^\text{Ht}_{i,j-1},\bm{X}_i,Z_i=z,D_i>j]$, where $\bm{\beta}^t_j(z)$ and $\bm{\beta}^{'t}_j(z)$ are defined in Assumption 5 and $Y^{Ht}_{i,j}$ is defined as $Y^{H}_{i,j}$ with $\bm{\beta}^t_j(z)$ and $\bm{\beta}^{'t}_j(z)$ in place of $\hat{\bm{\beta}}^{'}_j(z)$ and $\hat{\bm{\beta}}_j(z)$.
\end{lemma}

\begin{proof}
First, note that the definition of $Y^{Ht}_{i,j}$ and $E[Y_{i,j}\mid \overline{\bm{Y}}^{Ht}_{i,j-1},\bm{X}_i,Z_i=z,D_i>j]$ is such that $Y^{Ht}_{i,j}=R_{i,j}\mathbbm{1}(D_i> j)Y_{i,j}+\{1-R_{i,j}\mathbbm{1}(D_i> j)\}E[Y_{i,j}\mid \overline{\bm{Y}}^{Ht}_{i,j-1},\bm{X}_i,Z_i=z,D_i>j]$.

The proof now follows by induction. For the base case, note that $E[Y_{i,1}\mid Y^\text{obs}_{i,0},\bm{X}_i,Z_i=z,D_i>1]=\bm{\beta}^t_1(z)\cdot Y_{i,0}+\bm{\beta}^{'t}_1(z)\cdot (1,\bm{X}_i)$ if $R_{i,0}=1$ or $\bm{\beta}^t_1(z)\cdot E[Y_{i,0} \mid \bm{X_i},Z_i=z,D_i>0]+\bm{\beta}^{'t}_1(1)\cdot (1,\bm{X}_i)$ if $R_{i,0}=0$. As a single equation, this result can be expressed as $E[Y_{i,1}\mid Y^\text{obs}_{i,0},\bm{X}_i,Z_i=z,D_i>1]=\bm{\beta}^t_1(z)\cdot Y^{Ht}_{i,0} +\bm{\beta}^{'t}_1(z)\cdot (1,\bm{X}_i)$ since by definition $Y^{Ht}_{i,0}=R_{i,0}Y_{i,0}+(1-R_{i,0})E[Y_{i,0} \mid \bm{X_i},Z_i=z,D_i>0]$. Hence, the base case is true.

Now assume that the lemma is true for $j=0,\ldots,k-1$. We now prove that that the lemma must also be true for $j=k$. By Assumption 4, it is easy to show that 

\begin{equation}\label{res1}
E[Y_{i,k}\mid \overline{\bm{Y}}^\text{obs}_{i,k-1},\bm{X}_i,Z_i=z,D_i>k]=\bm{\beta}^t_k(z)\cdot \overline{E[Y_{i,k-1}\mid \overline{\bm{Y}}^\text{obs}_{i,k-1},\bm{X}_i,Z_i=z,D_i>k-1]}+\bm{\beta}^{'t}_k(z)\cdot (1,\bm{X}_i)
\end{equation}

Now, note that the $l$th element in $\overline{E[Y_{i,k-1}\mid \overline{\bm{Y}}^\text{obs}_{i,k-1},\bm{X}_i,Z_i=z,D_i>k-1]}$ can be represented by $R_{i,l}Y_{i,l}+(1-R_{i,l})E[Y_{i,l}\mid \overline{\bm{Y}}^\text{obs}_{i,l-1},\bm{X}_i,Z_i=z,D_i>l]$, which is equal to $Y^{Ht}_{i,l}=R_{i,l}Y_{i,l}+(1-R_{i,l})E[\overline{\bm{Y}}_{i,l}\mid \overline{\bm{Y}}^{Ht}_{i,l-1},\bm{X}_i,Z_i=z,D_i>l]$ by the lemma at $j=l$. Thus, $\overline{E[Y_{i,k-1}\mid \overline{\bm{Y}}^\text{obs}_{i,k-1},\bm{X}_i,Z_i=z,D_i>k-1]}$ can be replaced by $\overline{\bm{Y}}^{Ht}_{i,k-1}$ in \eqref{res1}, which upon doing so completes the proof. 
\end{proof}

\begin{lemma}\label{Lem2}
$\bm{\beta}^{'p}_j(z)=\bm{\beta}^{'t}_j(z)$ and $\bm{\beta}^{p}_j(z)=\bm{\beta}^{t}_j(z)$, where $\bm{\beta}^{'p}_j(z)$ and $\bm{\beta}^{p}_j(z)$ are the probability limits of $\hat{\bm{\beta}}^{'}_j(z)$ and $\hat{\bm{\beta}}_j(z)$, and $\bm{\beta}^{'t}_j(z)$ and $\bm{\beta}^{t}_j(z)$ are defined in Assumption 5. Additionally, $Y^{Hp}_{i,j}=Y^{Ht}_{i,j}$, where $Y^{Hp}_{i,j}$ is the probability limit of $Y^{H}_{i,j}$ and $Y^{Ht}_{i,j}$ is defined in Lemma 1.
\end{lemma}

\begin{proof}
Recall that $\hat{\bm{\beta}}^{'}_j(z)$ and $\hat{\bm{\beta}}_j(z)$ are obtained by solving the following estimating equations:
\begin{equation}\label{est_eq}
\begin{split}
\sum_{i=1}^nR_{i,j}\mathbbm{1}(Z_i= z)\mathbbm{1}(D_i> j)\{Y_{i,j}-\bm{\beta}^{'}_j(z)\cdot(1,\bm{X}_i)-\bm{\beta}_j(z)\cdot \overline{\bm{Y}}^H_{i,j-1}\}(1,\bm{X}_i,\overline{\bm{Y}}^H_{i,j-1})=0.
\end{split}
\end{equation}
The asymptotic bias of this estimating function is
\begin{align*}
&\left\{E[Y_{i,j}\mid \overline{\bm{Y}}^\text{obs}_{i,j-1},\bm{X}_i,Z_i=z,D_i> j, R_{i,j}=1]-\bm{\beta}^{'p}_j(z)\cdot(1,\bm{X}_i)-\bm{\beta}^p_j(z)\cdot \overline{\bm{Y}}^{Hp}_{i,j-1}\right\}(1,\bm{X}_i,\overline{\bm{Y}}^H_{i,j-1})\\
=&\left\{E[Y_{i,j}\mid \overline{\bm{Y}}^\text{obs}_{i,j-1},\bm{X}_i,Z_i=z,D_i> j]-\bm{\beta}^{'p}_j(z)\cdot(1,\bm{X}_i)-\bm{\beta}^p_j(z)\cdot \overline{\bm{Y}}^{Hp}_{i,j-1}\right\}(1,\bm{X}_i,\overline{\bm{Y}}^H_{i,j-1})\\
=&\left\{\bm{\beta}^{'t}_j(z)\cdot(1,\bm{X}_i)+\bm{\beta}^t_j(z)\cdot \overline{\bm{Y}}^{Ht}_{i,j-1}-\bm{\beta}^{'p}_j(z)\cdot(1,\bm{X}_i)-\bm{\beta}^p_j(z)\cdot \overline{\bm{Y}}^{Hp}_{i,j-1}\right\}(1,\bm{X}_i,\overline{\bm{Y}}^H_{i,j-1})
\end{align*}
where the second line follows from Assumptions 2  and 3, and the third line follows from Lemma~\ref{Lem1}.
We now prove the lemma by induction. First, consider the base case. Note that $\bm{\beta}^{'p}_0(z)=\bm{\beta}^{'t}_0(z)$ since the estimating function for $\bm{\beta}^{'}_0(z)$ is unbiased after imposing this equality. Next, we have by definition $Y^{Hp}_{i,0}=R_{i,0}\mathbbm{1}(D_i> 0)Y_{i,0}+\{1-R_{i,0}\mathbbm{1}(D_i> 0)\}\bm{\beta}^{'p}_0(z)\cdot(1,\bm{X}_i)$. As we have just shown that $\bm{\beta}^{'p}_0(z)=\bm{\beta}^{'t}_0(z)$, this implies that $Y^{Hp}_{i,0}=Y^{Ht}_{i,0}$. This completes the proof for the base case.

Now assume that the lemma is true for $j=0,\ldots,k-1$. The estimating function is asymptotically unbiased when $j=k$ if $\bm{\beta}^{'p}_k(z)=\bm{\beta}^{'t}_k(z)$ and $\bm{\beta}^{p}_k(z)=\bm{\beta}^{t}_k(z)$ since we have just assumed that $\overline{\bm{Y}}^{Hp}_{i,k-1}=\overline{\bm{Y}}^{Ht}_{i,k-1}$. Hence, these equalities must hold. Next, we have by definition $Y^{Hp}_{i,k}=R_{i,k}\mathbbm{1}(D_i> k)Y_{i,k}+\{1-R_{i,k}\mathbbm{1}(D_i> k)\}\{\bm{\beta}^{'p}_k(z),\bm{\beta}^p_k(z)\}\cdot(1,\bm{X}_i,\overline{\bm{Y}}^{Hp}_{i,k-1})$. As we have just shown that $\bm{\beta}^{'p}_k(z)=\bm{\beta}^{'t}_k(z)$ and $\bm{\beta}^{p}_k(z)=\bm{\beta}^{t}_k(z)$, and as $\overline{\bm{Y}}^{Hp}_{i,k-1}=\overline{\bm{Y}}^{Ht}_{i,k-1}$ by the inductive hypothesis, this implies that $Y^{Hp}_{i,k}=Y^{Ht}_{i,k}$. This completes the proof.
\end{proof}

\begin{lemma}\label{Lem3}
$E[Y^{Ht}_{i,j}\mid \bm{X}_i,Z_i=z]=E[Y_{i,j}(M+1,z)\mid \bm{X}_i]:=\mu_{i,j}(z)$, where $\mu_{i,j}(z)$ is defined in Assumption 5.
\end{lemma}

\begin{proof}
First, note that
\begin{equation}\label{res2}
E[Y^{Ht}_{i,j}\mid \bm{X}_i,Z_i=z]=E[R_{i,j}\mathbbm{1}(D_i> j)Y_{i,j}+\{1-R_{i,j}\mathbbm{1}(D_i> j)\}\{\bm{\beta}^{'t}_j(z),\bm{\beta}^t_j(z)\}\cdot(1,\bm{X}_i,\overline{\bm{Y}}^{Ht}_{i,j-1}) \mid X_i, Z_i=z].
\end{equation}
We now prove the lemma by induction. Consider first the base case. Note that $E[Y_{i,0}(M+1,z)\mid \bm{X}_i,Z_i=z]=\bm{\beta}^{'t}_0(Z_i)\cdot(1,\bm{X}_i)$ by Assumption 4. It then follows that if $D_i=0$, \eqref{res2} simplifies to $E[Y^{Ht}_{i,0}\mid \bm{X}_i,Z_i=z]=E[Y_{i,0}(M+1,z)\mid \bm{X}_i]$. If however $D_i>0$, then \eqref{res2} simplifies to 
$E[Y^{Ht}_{i,j}\mid \bm{X}_i,Z_i=z]=E[R_{i,0}Y_{i0}+(1-R_{i,0})Y_{i,0}(M+1,z)\mid \bm{X}_i,Z_i=z]=E[Y_{i,0}(M+1,z)\mid \bm{X}_i]$ by the consistency assumption. Thus, in both cases the base case holds.

Now assume that the lemma is true for $j=0,\ldots,k-1$. Then, 
\begin{align*}
E\left[\{\bm{\beta}^{'t}_{k}(z),\bm{\beta}^t_{k}(z)\}\cdot (1,\bm{X}_i,\overline{\bm{Y}}^{Ht}_{i,k-1}) \mid X_i, Z_i=z\right]=&\{\bm{\beta}^{'t}_{k}(z),\bm{\beta}^t_{k}(z)\}\cdot\left(1,\bm{X}_i,E[\overline{\bm{Y}}_{i,k-1}(M+1,z) \mid \bm{X}_i]\right)\\
=&E[Y_{i,k}(M+1,z)\mid \bm{X}_i].
\end{align*}
By using this result and applying exactly the same argument as for the base case, i.e. considering the case when $D_i\leq k$ and $D>k$ separately, it is easy to show that the lemma also holds for $j=k$.
\end{proof}

Note that by Lemma \eqref{Lem2} and \eqref{Lem3}, $E[Y^{Hp}_{i,j}|\bm{X}_i,Z_i=z]=E[Y^{Ht}_{i,j}|\bm{X}_i,Z_i=z]=E[Y_{i,j}(M+1,z)\mid \bm{X}_i]$. It is also easy to show that this equation holds if $\bm{X}_i$ is replaced by $\bm{V}_i$. Hence, our proposed estimator for the hypothetical estimand is asymptotically unbiased. 

We are now ready to show that $E[Y^{CIRp}_{i,j}\mid \bm{X}_i,Z_i=1]=E[Y_{i,j}\mid \bm{X}_i,Z_i=1]$, from which it trivially follows that $E[Y^{CIRp}_{i,j}\mid \bm{V}_i,Z_i=1]=E[Y_{i,j}\mid \bm{V}_i,Z_i=1]$ upon integrating out the components of $\bm{X}_i$ that are not contained in $\bm{V}_i$. 

Recall that when $j < D_i$, $Y^{CIR}_{i,j}=Y^H_{i,j}$. Thus by Lemma~\eqref{Lem3} and the consistency assumption, $E[Y^{CIRp}_{i,j} \mid \bm{X}_i, Z_i=1]=E[Y_{i,j}\mid \bm{X}_i,Z_i=1]$ when $j < D_i$. We now show that $E[Y^{CIRp}_{i,j} \mid \bm{X}_i, Z_i=1]=E[Y_{i,j}\mid \bm{X}_i,Z_i=1]$ when $j\geq D_i=k$. To do this, we will make use of the following result when $j \geq k$.
\begin{align*}
Y^{CIRp}_{i,j}&=R_{i,j}Y_{i,j}+(1-R_{i,j})\left\{\bm{\beta}^p_{j}(0)\{\overline{\bm{Y}}^{CIRp}_{i,j-1}-\bm{\mu}^{CIRp}_{i,j-1}\}+\{\mu^p_{i,k-1}(1)-\mu^p_{i,k-1}(0)\}+\mu^p_{i,j}(0)\right\}\\
&=R_{i,j}Y_{i,j}+(1-R_{i,j})\left\{\bm{\beta}^t_{j}(0)\{\overline{\bm{Y}}^{CIRp}_{i,j-1}-\bm{\mu}^{CIR}_{i,j-1}\}+\{\mu_{i,k-1}(1)-\mu_{i,k-1}(0)\}+\mu_{i,j}(0)\right\}
\end{align*}
where the first line follows from the proposed definition of $Y^{CIR}_{i,j}$ upon replacing $\hat{\mu}_{i,j}(z)$ and $\hat{\bm{\mu}}^{CIR}_{i,j}$ by their probability limits $\mu^p_{i,j}(z)$ and $\bm{\mu}^{CIRp}_{i,j}$, and the second line follows from Lemmas~\eqref{Lem2} and~\eqref{Lem3}. The proof can now be completed by induction. First, consider the base case $j=k$. Note that if $R_{i,k}=1$, $Y^{CIRp}_{i,k}=Y_{i,k}$ and so it trivially follows that $E[Y^{CIRp}_{i,k} \mid \bm{X}_i, Z_i=1]=E[Y_{i,k}\mid \bm{X}_i,Z_i=1]$. If however $R_{i,k}=0$, then
\begin{align*}
E[Y^{CIRp}_{i,k} \mid \bm{X}_i, Z_i=1]=&E\left[\bm{\beta}^t_{j}(0)\{\overline{\bm{Y}}^{CIRp}_{i,k-1}-\bm{\mu}^{CIR}_{i,k-1}\}+\{\mu_{i,k-1}(1)-\mu_{i,k-1}(0)\}+\mu_{i,j}(0)\mid \bm{X}_i, Z_i=1\right]\\
=&\bm{\beta}^t_{j}(0)\{E[\overline{\bm{Y}}^{CIRp}_{i,k-1}\mid \bm{X}_i, Z_i=1]-\bm{\mu}^{CIR}_{i,k-1}\}+\{\mu_{i,k-1}(1)-\mu_{i,k-1}(0)\}+\mu_{i,j}(0)\\
=&\bm{\beta}^t_{j}(0)\{E[\overline{\bm{Y}}_{i,k-1}\mid \bm{X}_i, Z_i=1]-\bm{\mu}^{CIR}_{i,k-1}\}+\{\mu_{i,k-1}(1)-\mu_{i,k-1}(0)\}+\mu_{i,j}(0)\\
=&E\left[\bm{\beta}^t_{j}(0)\{\overline{\bm{Y}}_{i,k-1}-\bm{\mu}^{CIR}_{i,k-1}\}+\{\mu_{i,k-1}(1)-\mu_{i,k-1}(0)\}+\mu_{i,j}(0)\mid \bm{X}_i, Z_i=1\right]\\
=&E\left[E[Y_{i,k} \mid \overline{\bm{Y}}_{i,k-1},\bm{X}_i, Z_i=1,D_i=k]\mid\bm{X}_i, Z_i=1\right]\\
=&E\left[Y_{i,k} \mid\bm{X}_i, Z_i=1\right].
\end{align*} 

Here, the fifth line follows from Assumption 5. This completes the base case. Upon assuming the inductive hypothesis for $l=0,\ldots j-1$, i.e., $E[Y^{CIRp}_{i,l} \mid \bm{X}_i, Z_i=1]=E[Y_{i,l}\mid \bm{X}_i,Z_i=1]$, it is easy to show using exactly the same arguments as for the base case that the sequential definition of $Y^{CIRp}_{i,j}$ implies $E[Y^{CIRp}_{i,j} \mid \bm{X}_i, Z_i=1]=E[Y_{i,j}\mid \bm{X}_i,Z_i=1]$.

In summary, we have shown that our estimator of the treatment policy estimand under CIR is asymptotically unbiased.

\section{Inference}
Following \cite{Wolbers2022}, we propose to use re-sampling based methods, e.g. the jackknife and non-parametric bootstrap, to make inference, e.g. to estimate confidence intervals. This is justified because our estimator is an M-estimator as it only involves solving estimating equations from linear models, and \cite{Shao1995} prove that the jackknife (see Example 2.5 in Section 2 on page 37-38) and bootstrap (see Theorem 3.6 in Section 3 on page 83-84) are valid inferential methods for M-estimators. 

The jackknife method to estimate confidence intervals involves applying the following steps. First, create $n$ data sets labelled $\mathcal{D}_i$ by removing the $i$th patient's data from the observed data set. Second, apply the proposed methods to each data set $\mathcal{D}_i$ $(i=1,\ldots,n)$ to obtain an estimate of the treatment effect of interest. Let $\hat{\theta}_i$ be the treatment effect estimate based on $\mathcal{D}_i$, and $\overline{\hat{\theta}}$ be the sample mean of $\hat{\theta}_i$ $(i=1,\ldots,n)$. Third, define $$SE_{jack}=\sqrt{\frac{n-1}{n}\sum_{i=1}^n(\hat{\theta}_i-\overline{\hat{\theta}})^2}.$$ Finally, compute $100(1-\alpha)\%$ confidence intervals as $\hat{\theta}\pm z_{1-\alpha/2} SE_{jack}$, where $\hat{\theta}$ is the treatment effect estimate based on the observed data, i.e. without removing any patient's data, and $z_{1-\alpha/2}$ is the $100(1-\alpha/2)$ percentile of the standard normal distribution. In contrast, the non-parametric bootstrap method to create confidence intervals involves applying the following steps. First, create $m$ data sets labelled $\mathcal{D}_i$ by re-sampling patients with replacement in each treatment arm and within each level of the randomization stratification factors until $\mathcal{D}_i$ contain the same number of patients in each strata of each treatment arm. Second, apply the proposed methods to each data set $\mathcal{D}_i$ $(i=1,\ldots,m)$ to obtain an estimate of the treatment effect of interest. Again, let $\hat{\theta}_i$ be the treatment effect estimate based on $\mathcal{D}_i$. Finally, compute $100(1-\alpha)\%$ confidence intervals as either $(\hat{\theta}_{[100\alpha/2\%]},\hat{\theta}_{[100(1-\alpha/2)\%]})$ (percentile approach), where $\hat{\theta}_{[q\%]}$ is the $q$th percentile of $\hat{\theta}_i$, or $\hat{\theta}\pm z_{1-\alpha/2} sd(\hat{\theta}_i)$ (normal approximation approach), where $sd(\hat{\theta}_i)$ is the sample standard deviation of $\hat{\theta}_i$. In this article, we focus on the percentile approach because the normal approximation approach gives almost identical results to the jackknife. The choice of $m$, i.e. the number of bootstrap samples, could be based on pragmatic considerations and achieving negligible Monte Carlo error with respect to the chosen level of precision for reporting results.  

The percentile approach to the bootstrap may be preferable to the jackknife if it is believed that a normal distribution does not provide a good approximation to the distribution of the proposed estimator, e.g. due to the sample size being small. In contrast, the jackknife may be preferable when Monte Carlo error is a legitimate concern, e.g. because it is not pragmatic to reduce Monte Carlo error by increasing the number of bootstrap samples, and/or when non-convergence in bootstrap samples is probable, e.g. due to some levels of discrete baseline covariates being rarely observed. In Section~\ref{Sim_sec}, we assess the performance of the bootstrap and jackknife for making inference with our proposed estimator.   

\section{Comparison between implementing MMRM and SLR}\label{MMRMvsSLR}
In this section, we contrast the implementation of MMRM and SLR, and provide heuristic arguments for why CMI with SLR can be much more computationally efficient and less prone to convergence issues than with MMRM. 

Recall that an MMRM is a multivariate normal model characterized by parameters for the mean and variance covariance matrix. Assuming that there are no covariates, the MMRM model contains $M$ parameters for the mean and $M(M+1)/2$ parameters for an unstructured variance covariance matrix, where $M$ is the intended number of outcomes per patient. In contrast, the SLR approach requires factorizing the multivariate normal distribution into a sequence of condition normal models for the outcome at each visit conditional on the previous outcomes, and then estimating the conditional means (but not the variance). Therefore, the mean model for the $j$th outcome will contain $j$ parameters (i.e. the intercept term and regression coefficients for the $j-1$ previous outcomes), which in total implies that there are $M(M+1)/2$ parameters to estimate when there are $M$ intended outcomes. In conclusion, the implementation of SLR requires the estimation of $M$ fewer parameters than MMRM when there are $M$ intended outcomes per patient.

Furthermore, recall that the parameters of the MMRM are estimated simultaneously, whereas the SLR approach estimates the parameters for the conditional mean outcomes sequentially. The advantage of the sequential approach is that it can be implemented by fitting $M$ distinct linear models, and there are closed form solutions for the regression coefficients of linear models, e.g. as stated in Section \ref{Prop_est}. In contrast, the fitting of MMRM requires an iterative procedure to search for the optimal parameter values, e.g. the ‘Broyden–Fletcher–Goldfarb–Shannon’ optimization technique \cite{Broyden1970} to search for the maximum likelihood estimates, because a closed form solution is difficult to obtain.

Overall, the implementation of SLR requires fewer parameters to be estimated compared to MMRM and does not require an iterative procedure to search for parameter estimates, unlike MMRM. Consequently, SLR can be much faster to implement and be less prone to convergence issues, particularly when MMRM contain many parameters and many iterations of the search algorithm are required to obtain parameter estimates (e.g. the initial values are very far from the optimal ones or if the likelihood function is relatively flat). However, because the exact difference in computation time of the CMI with MMRM and SLR is difficult to quantify due to their dependence on the precise algorithm used for parameter estimation and the data, we use simulations and a real data example in the subsequent sections to illustrate the practical differences of these approaches.

\section{Simulations}\label{Sim_sec}
In this section, we performed simulations to verify our theoretical results in Section~\ref{Jus_sec}, i.e. our proposed estimator converges in probability to the same limit as the complete data estimator, and to assess the performance of the bootstrap and jackknife for making inference. The R code for the simulations can be found in the Supplementary Materials.
\subsection{Data generating mechanism}
The data generating mechanism was designed using data from the CREAD study in Alzheimer's disease \cite{Ostrowitzki2022}. Specifically, the distribution of the outcomes at each visit, the randomization stratification variables, and the relationship between these variables in the control arm of the CREAD study were used to select parameters for the data generating mechanisms of the simulation studies.

We assigned 500 patients to receive the active treatment and 500 patients to receive the control treatment. For each patient in each treatment arm, we simulated three independent binary baseline covariates $X_1$, $X_2$ and $X_3$ from Bernoulli distributions with success probabilities of 0.7 for $X_1$ and $X_2$ and 0.4 for $X_3$. The baseline outcome $Y_0$ was simulated from a normal distribution with a mean of 3.84 and standard deviation of 1.64. We then simulated the partially observed always treated longitudinal outcomes for visits 1-5, $\overline{\bm{Y}}_j(6,z)$ $(j=1,\ldots,5)$, from a multivariate normal distribution with mean vector $\bm{\lambda}_{0}+0.03Y_0-0.02X_1+0.45X_2-0.82X_3$ and covariance matrix $$\Sigma=\begin{pmatrix}
 4.28 & 4.02 & 4.29 & 4.58 & 4.73\\
4.02 & 8.41 &  7.87 &  8.13 &  8.22\\
4.29 & 7.87 & 14.21 & 13.97 & 13.87\\
4.58 & 8.13 & 13.97 & 20.43 & 20.44\\
4.73 & 8.22 & 13.87 & 20.44 & 24.70
\end{pmatrix},$$    
where $\bm{\lambda}_{0}=(0.41,1.29,2.17,3.33,4.05)$ for patients in the control arm and $\bm{\lambda}_{0}=(0.41,1.22,1.83,2.55,3.10)$ for patients in the active arm. In order to generate a null treatment effect, $\bm{\lambda}_{0}$ in the active arm was also set equal to $\bm{\lambda}_{0}$ in the control arm. These parameter values result in the true value of the hypothetical estimand at the final visit (i.e. $j=5$) being equal to -0.95 in the non-null treatment effect setting. Next, we set the probability of remaining on the assigned treatment at visits $j>1$ conditional on still being on the assigned treatment at the previous visit to $$\frac{1}{1+\exp(-2.75+0.04Y_{j-1}(6,z)+0.01Y_0)}.$$ 
Based on a large number of simulations, this results in the true value of the treatment policy estimand at the final visit being equal to -0.79 under CIR and -0.72 under J2R in the non-null treatment effect setting. Finally, we introduced two sources of missing data. The first results in patients withdrawing from study with probability 0.75 as soon as they discontinue their assigned treatment. The second consists of assigning each observation with a probability of 0.05 of being set to missing. Overall, this results in approximately 12.3$\%$ of the data being set to missing in all settings. 
\subsection{Results}
Table~\ref{sim_results} displays the results from estimating the treatment policy estimand at the final visit in 1000 simulated data sets using the following methods: 1) complete data estimator defined as the estimated regression coefficient of the treatment assignment from regressing the complete outcome data (i.e. before missing data sources are introduced) on the treatment assignment, and baseline covariates and outcome, 2) CMI based on the proposed SLR approach in Section~\ref{Prop_est} and 3) CMI based on MMRM for the imputation model with the interactions between (categorical) visit and all baseline covariates and treatment assignment in the mean functions and a common unstructured covariance matrix across treatment arms, as described in Section~\ref{MMRM_sec}. 95$\%$ confidence intervals for the complete data estimator were based on the Wald approach. CMI based on MMRM was included in the simulations to provide context for the performance of the proposed estimator of the treatment effect. However, it was too computationally intensive to implement the bootstrap with 1000 samples and jackknife for the CMI with MMRM approach. For instance, in the non-null treatment effect setting where data are generated under CIR, it took approximately 272.3 and 246.5 minutes to implement the bootstrap and jackknife for one simulated data on a single core laptop with an Intel Core i5-1145G7 processor (2.60 GHz), 16 GB RAM, and Windows 10 Enterprise. In contrast, the bootstrap and jackknife took approximately 15.5 and 13.1 minutes to implement in the same setting with the CMI with SLR approach.

\begin{table}[h!]
\center
\caption{Simulation results for the treatment policy estimand at the final visit from the complete data estimator, CMI with the proposed SLR approach, and CMI with MMRM.}
\begin{tabular}{ |c|c|c|c|c|c| }
 \hline
Method & Bias & RMSE  & Coverage probability & Power & Type I error   \\
\hline
\multicolumn{6}{|c|}{\textbf{Simulation setting 1:} Data generated under CIR assumption with non-null treatment effect}\\
\hline
Complete data estimator &-0.009 &0.314 &94.8$\%$ &72.5$\%$ &- \\
CMI based on SLR with bootstrap &-0.001 &0.301 &94.4$\%$ &76.3$\%$ &- \\
CMI based on SLR with jackknife &-0.001 &0.301 &94.7$\%$ &75.4$\%$ &- \\
CMI based on MMRM &-0.001 &0.300 &- &- &- \\
\hline
\multicolumn{6}{|c|}{\textbf{Simulation setting 2:} Data generated under CIR assumption with no treatment effect}\\
\hline
Complete data estimator &0.016 &0.313 &95.2$\%$ &- &4.8$\%$ \\
CMI based on SLR with bootstrap &0.015 &0.308 &94.1$\%$ &- &5.9$\%$ \\
CMI based on SLR with jackknife &0.015 &0.308 &94.9$\%$ &- &5.1$\%$ \\
CMI based on MMRM &0.016 &0.307 &- &- &- \\
\hline
\multicolumn{6}{|c|}{\textbf{Simulation setting 3:} Data generated under J2R assumption with non-null treatment effect}\\
\hline
Complete data estimator &0.005 &0.306 &95.6$\%$ &62.5$\%$ &- \\
CMI based on SLR with bootstrap &0.008 &0.268 &95.5$\%$ &74.0$\%$ &- \\
CMI based on SLR with jackknife &0.008 &0.268 &95.7$\%$ &73.8$\%$ &- \\
CMI based on MMRM &0.008 &0.268 &- &- &- \\
\hline
\multicolumn{6}{|c|}{\textbf{Simulation setting 4:} Data generated under J2R assumption with no treatment effect}\\
\hline
Complete data estimator &-0.007 &0.310 &94.9$\%$ &- &5.1$\%$ \\
CMI based on SLR with bootstrap &-0.008 &0.264 &95.2$\%$ &- &4.8$\%$ \\
CMI based on SLR with jackknife &-0.008 & 0.264&95.6$\%$ &- &4.4$\%$  \\
CMI based on MMRM &-0.007 &0.264 &- &- &- \\
\hline
\end{tabular}
\label{sim_results}
\footnotesize{RMSE: root mean squared error; CIR: copy increments in reference; CMI: conditional mean imputation; SLR: sequential linear regression; MMRM: mixed models for repeated measures; J2R: jump to reference.}\\
\end{table}
From the table, it can be seen that the proposed SLR approach provides similar treatment effect estimates to the other estimators, as expected by the theory in Section~\ref{Jus_sec}, and all estimators exhibit negligible bias. However, the SLR and MMRM imputation approaches have slightly smaller RMSE than the complete data estimator. This occurred because replacing missing data with a conditional mean estimate under reference based assumptions reduced the variation in the treatment effect estimates in the simulation study, which also has the consequence of increasing statistical power to detect treatment effects. More generally, it has been shown in simulations that estimators relying on reference based assumptions can become increasingly more precise as the amount of missing data post treatment discontinuation increases \cite{Bartlett2023}. These observations further reinforce the recommendation to only consider RBI methods if the reference based assumption is likely to be accurate and the proportion of missing data post treatment discontinuation is low. Regarding coverage probabilities of confidence intervals, the bootstrap and jackknife achieved close to nominal coverage, and performed similarly to the conventional Wald approach in the absence of missing data. Consequently, both methods controlled type I error reasonably well. Finally, it is worth noting that no convergence issues were observed with the proposed estimator in any of the simulated data sets including in bootstrap and jackknife samples.

Overall, the simulations confirmed that the proposed estimator can provide unbiased treatment effect estimates and confidence intervals with nominal coverage in the presence of missing data, is unlikely to result in convergence issues, and can be less computationally intensive to implement than CMI with MMRM.

\section{Application}
We applied the proposed CMI with SLR approach and its counterpart based on MMRM to a publicly available example data set from an antidepressant clinical trial of an active drug versus placebo. This data set can be found in the rbmi package in R \cite{Gower-Page2022}. The endpoint of interest is the Hamilton 17-item rating scale for depression (HAMD17) which was assessed at baseline and at weeks 1, 2, 4, and 6. Higher values of the score indicates worse outcomes. Discontinuation of study drug occurred in 24$\%$ (20/84) and 26$\%$ (23/88) of patients in the active and placebo arm respectively; see Table~\ref{Discon_table} for more details. All data after study drug discontinuation were missing and there was a single missing observation prior to study drug discontinuation. The imputation model for the CMI with MMRM approach had the change from baseline in the HAMD17 score as the outcome, and the covariates in the mean function included the main effects and all interactions between the treatment group, (categorical) visit label and baseline HAMD17 score, and a common unstructured covariance matrix in both groups was assumed. The mean HAMD17 score at baseline was 18.6 (SD=5.9) and 17.2 (SD=5.1) in the active and placebo group, respectively. For the CMI with SLR approach, the imputation model for outcomes in each treatment arm at each visit included all previous outcomes and the baseline HAMD17 score. That is, unlike the CMI with MMRM approach, the CMI with SLR approach does not assume that the correlations between longitudinal outcomes are the same across treatment arms. Finally, the analysis model for the mean change in outcome at the final visit included the treatment group and the baseline HAMD17 score. A short tutorial providing detailed descriptions on the data set and how the proposed approach can be applied to the data set can be found in the Supplementary Materials. 

\begin{table}[ht!]
\center
\caption{Number of patients that discontinued study drug up to and including each visit in the antidepressant clinical trial}
\begin{tabular}{ |c|c|c|c|c|c| }
 \hline
 \multirow{2}{5em}{Treatment group} & \multicolumn{5}{c|}{Number ($\%$) of patients that discontinued study drug by each visit} \\ 
 \cline{2-6}
&\hspace*{1.5mm} Baseline \hspace*{1.5mm}&\hspace*{1.5mm} Week 1 \hspace*{1.5mm}&\hspace*{1.5mm} Week 2 \hspace*{1.5mm}&\hspace*{1.5mm} Week 4 \hspace*{1,5mm}&\hspace*{1.5mm} Week 6\hspace*{1.5mm}\\
\hline
Active &0 (0$\%$) &0 (0$\%$)&6 (7.1$\%$) &11 ($13.1\%$)&20 (23.8$\%$) \\
Placebo &0 (0$\%$) &0 (0$\%$) &7 (8.0$\%$) &12 ($13.6\%$) &23 (26.1$\%$) \\
\hline
\end{tabular}
\label{Discon_table}
\end{table}

The results for CMI with MMRM and SLR using jackknife and bootstrap with 1000 bootstrap samples are reported in Table~\ref{Tab_results}. From the table, it can be seen that both approaches provide similar results in terms of the marginal mean estimates in each treatment arm at the final visit as well as the treatment effect estimates under both J2R and CIR. Unsurprisingly, the treatment effect estimates were larger when assuming that the treatment effect at the last visit prior to treatment discontinuation was retained for all subsequent visits, i.e. the CIR assumption, compared to assuming that the treatment effect was completely lost with treatment discontinuation, i.e. the J2R assumption. In any case, both treatment effect estimates were statistically significant, and the estimates of the mean change in HAMD17 outcomes at the final visit in the active but not control arm were clinically substantial, since according to \cite{Rush2021} changes of 7-12 points in HAMD17 can be regarded as clinically substantial based on anchored based analyses. 

Regarding the confidence intervals for the treatment effect, the jackknife resulted in slightly wider intervals when applying CMI with SLR compared to MMRM, but the bootstrap resulted in slightly wider intervals when applying CMI with MMRM compared to SLR. For both methods, wider intervals were observed with the jackknife compared to the bootstrap, and wider intervals were observed when CIR was assumed compared to when J2R was assumed. The observation of narrower confidence intervals with the bootstrap is consistent with the simulation study where slightly higher power was observed with the bootstrap. The observation of wider confidence intervals under CIR is expected because CMI under CIR depends on more unknown quantities than J2R as it additionally encodes the unknown treatment effect at the last visit prior to treatment discontinuation. The computing time on a single core laptop with an Intel Core i5-1145G7 processor (2.60 GHz), 16 GB RAM, and Windows 10 Enterprise to obtain both reported treatment effect estimates and 95$\%$ confidence intervals were 69.90 and 11.23 minutes for CMI based on MMRM with bootstrap and jackknife, respectively, and 4.76 and 0.79 minutes for CMI based on SLR with bootstrap and jackknife, respectively.

Overall, the proposed CMI with SLR approach was able to provide similar results to the recently proposed CMI with MMRM approach, despite requiring significantly less computation time. In terms of the results, both approaches suggested that a strong and clinically meaningful treatment effect estimate would have been observed in the absence of missing data even if the treatment effect is only partially retained (CIR assumption) or even completely lost after treatment discontinuation (J2R assumption).

\begin{table}[ht!]
\center
\caption{Parameter estimates and 95$\%$ confidence intervals for the marginal means and treatment policy estimand at the final visit from CMI based on SLR and MMRM.}
\begin{tabular}{ |c|c|c|c| }
 \hline
 \multirow{2}{4em}{Estimators} & \multicolumn{2}{c|}{Marginal mean of outcomes at the final visit} & \multirow{2}{7.5em}{Treatment effect} \\ 
 \cline{2-3}
& Active arm & Control arm &  \\
\hline
\multicolumn{4}{|c|}{\textbf{Estimators relying on the CIR assumption}}\\
\hline
CMI based on SLR with bootstrap &-7.480 (-9.153 -6.015) &-4.614 (-6.146 -3.175) &-2.453 (-4.464 -0.567) \\
CMI based on SLR with jackknife &-7.480 (-8.943 -6.017) &-4.614 (-6.213 -3.015) &-2.453 (-4.449 -0.458) \\
CMI based on MMRM with bootstrap &-7.477 (-9.118 -5.870) &-4.602 (-5.960 -3.198) &-2.464 (-4.393 -0.488) \\
CMI based on MMRM with jackknife &-7.477 (-9.072 -5.883) &-4.602 (-6.054 -3.150) &-2.464 (-4.443 -0.485) \\
\hline
\multicolumn{4}{|c|}{\textbf{Estimators relying on the J2R assumption}}\\
\hline
CMI based on SLR with bootstrap &-7.177 (-8.536 -5.758) &-4.614 (-5.966 -3.149) &-2.179 (-3.903 -0.575) \\
CMI based on SLR with jackknife &-7.177 (-8.640 -5.714) &-4.614 (-6.049 -3.179) &-2.179 (-3.909 -0.449) \\
CMI based on MMRM with bootstrap &-7.176 (-8.639 -5.782) &-4.602 (-6.108 -3.225) &-2.192 (-3.967 -0.565) \\
CMI based on MMRM with jackknife &-7.176 (-8.610 -5.743) &-4.602 (-6.054 -3.150) &-2.192 (-3.906 -0.478)  \\
\hline
\end{tabular}
\label{Tab_results}
\footnotesize{CIR: copy increments in reference; CMI: conditional mean imputation; SLR: sequential linear regression; MMRM: mixed models for repeated measures; J2R: jump to reference.}\
\end{table}

\section{Discussion}
In this article, we proposed and justified the use of SLR for CMI of longitudinal continuous outcomes under reference based assumptions. In particular, our simulations confirmed that the proposed approach provides similar results to the complete data estimator on average across simulations, as implied by our theoretical results, and almost identical results to CMI with MMRM in each simulated data set and the real data example. The main motivation for developing the SLR approach was to decrease the computation time of implementing the recently proposed CMI with MMRM and the jackknife or bootstrap for inference. For instance, CMI with MMRM required over 4 hours to implement when the jackknife or bootstrap was applied to one of our simulated data set of 1000 patients with 5 planned follow-up visits after baseline. In contrast, the proposed SLR approach could be implemented in approximately 15 minutes when applied to the same data sets. The substantial gain in computation time with SLR can be mainly attributed to decoupling of the estimation of parameters indexing the imputation models at each visit. This also has other advantages such as greater flexibility to specify dependence structures between longitudinal outcomes, since most MMRM software do not allow user-specified covariance matrices, and useful extensions of linear models can be incorporated into the CMI procedure, e.g. penalized regression or other approaches to reduce the influence of outliers \cite{Liu2022}. Providing more flexibility to specify arbitrary outcome dependence structures is particularly attractive when there are many visits and it is desirable to restrict the dependence of each outcome to a limited number of previous outcomes without being overly simplistic. However, the disadvantage of the proposed approach is that it is no longer possible to share parameters across imputation models for outcomes at each visit. Therefore, it may not be able to fully benefit from a more parsimonious specification of the imputation models when such a specification is reasonable.        

Regarding directions of future research, it would be useful to develop an approach to obtain information anchored inference in the CMI framework, since such inferences can be used to complement the standard frequentist inferences that are proposed in this article. A potential solution would be to use the standard errors obtained for the estimators of the hypothetical estimand, which do not rely on reference based assumptions, to prevent these assumptions from increasing the preciseness of estimators of the treatment policy estimand. However, the performance of this approach would need to be evaluated in simulations to demonstrate its usefulness. Second, it would be important to extend the proposed approach to other outcome types. This is in line with recent developments in the multiple imputation framework, e.g. see \cite{Fang2022} and \cite{Atkinson2019} for proposed approaches to implement multiple imputation under reference based assumptions for binary and time-to-event outcomes. Third, if the proposed approach is also required for observational studies, then it will need to be adapted to the irregular visit times setting and to be able to handle a large number of visits. As mentioned previously, the proposed SLR approach has been developed by specifying separate parameters for the imputation models at each visit. However, with a large number of visits, the number of parameters to estimate could become large and this could introduce a lot of uncertainty into treatment effect estimation.  Therefore, further developing the proposed approach with more parsimonious specifications of the imputation models that allow parameters to be shared across visits would greatly help alleviate this problem for observational settings. Finally, it would be worthwhile to explore how incompletely observed time-varying covariates can be incorporated into the imputation models, particularly when it is known that such covariates are highly predictive of the outcomes.  These and other research topics are currently being investigated by us.

\section*{Acknowledgements}
The author would like to thank Dr Marcel Wolbers for providing constructive feedback that greatly improved this article.

\section*{Competing Interests}
The author has no competing interests to declare that are relevant to the content of this article.

\end{document}